\documentclass[american, aps,prd, twocolumn, superscriptaddress, reprint]{revtex4-1}
\newcommand{\be}{\begin{eqnarray}}
\newcommand{\ee}{\end{eqnarray}}

\def\numu{{\nu_{\mu}}}

\newcommand{\ms}{\Delta m^2_{21}}
\newcommand{\ma}{\Delta m^2_{31}}

\newcommand{\sss}{\sin^2 \theta_{12}}

\newcommand{\dcp}{\delta_{CP}}

\def\nn{\nonumber}

\def\gs{\mathrel{
   \rlap{\raise 0.511ex \hbox{$>$}}{\lower 0.511ex \hbox{$\sim$}}}}
\def\ls{\mathrel{
   \rlap{\raise 0.511ex \hbox{$<$}}{\lower 0.511ex \hbox{$\sim$}}}}
\newcommand{\bea}{\begin{equation} \begin{array}{c}}
\newcommand{\bead}{\begin{equation} \begin{array}{cccc}}
\newcommand{\eea}{ \end{array} \end{equation}}

\usepackage[unicode=true,pdfusetitle, bookmarks=true,bookmarksnumbered=false,bookmarksopen=false, breaklinks=false,pdfborder={0 0 0},backref=false,colorlinks=false] {hyperref}
\hypersetup{ colorlinks,linkcolor=myurlcolor,citecolor=myurlcolor,urlcolor=myurlcolor}
\usepackage{braket,colortbl,amsthm,amsmath,amssymb,txfonts,cleveref}
\definecolor{myurlcolor}{rgb}{0,0,0.7}
\usepackage{graphics,graphicx}
\usepackage{color}
\usepackage{graphicx}
\usepackage[format = plain,labelfont = bf,up, textfont = normal , up, justification =raggedright, singlelinecheck =false]{caption}
\usepackage{subcaption}
\usepackage{tabularx}
\usepackage{amsmath}

\usepackage{graphicx}
\usepackage[utf8x]{inputenc}
\usepackage{color}
\usepackage{amsmath}
\usepackage{braket}
\usepackage{latexsym}
\usepackage{amssymb}
\usepackage{amsthm}
\usepackage{bm}
\usepackage{graphics,epstopdf}
\usepackage{color}\usepackage{amsmath}
\usepackage{enumitem}
\usepackage{fmtcount}
\usepackage{booktabs}
\usepackage{epsfig}
\def\slc#1{\setbox0=\hbox{$#1$}           
    \dimen0=\wd0                                 
    \setbox1=\hbox{/} \dimen1=\wd1               
    \ifdim\dimen0>\dimen1                        
       \rlap{\hbox to \dimen0{\hfil/\hfil}}      
       #1                                        
    \else                                        
       \rlap{\hbox to \dimen1{\hfil$#1$\hfil}}   
       /                                         
    \fi}

\def\numu{{\nu_{\mu}}}

\begin{document}

\title{On Resolving the Dark LMA Solution at Neutrino Oscillation Experiments}

\author{Sandhya Choubey}
\email{choubey@kth.se}
\affiliation{Department of Physics, School of Engineering Sciences,\\
KTH Royal Institute of Technology, AlbaNova University Center,\\
Roslagstullsbacken 21, SE-106 91, Stockholm, Sweden}

\author{Dipyaman Pramanik}
\email{dipyaman@prl.res.in}
\affiliation{Theoretical Physics Division, Physical Research Laboratory,\\ Navrangpura, Ahmedabad - 380009, India}

\begin{abstract}
In presence of non standard interactions, the solar neutrino data is consistent with two solutions, one close to the standard LMA solution with $\sin^2\theta_{12} \simeq 0.3$ and another with $\sin^2\theta_{12}^D \simeq 0.7(=\cos^2\theta_{12})$. The latter has been called the Dark LMA (DLMA) solution in the literature. This issue is hard to resolve via oscillations because of the existence of the so-called ``generalised mass hierarchy'' degeneracy of the neutrino mass matrix in presence of NSI. 
However, if the mass hierarchy is independently determined in a non-oscillation experiment such as neutrino-less double beta decay, the invariance of neutrino oscillation probabilities under $\sin^2\theta_{12} \leftrightarrow \cos^2\theta_{12}$ is lost and the possibility of resolving the LMA vs DLMA opens up. We point out that the $P_{\mu\mu}$ channel can distinguish $\theta_{12}$ from $\theta_{12}^D$ and study the corresponding difference in long-baseline experiments. We show that a key ingredient required is the input from the $P_{ee}$ channel measured at a reactor experiment.  We find that if the mass hierarchy is determined by neutrino-less double beta decay, then a combined measurement of the effective mass squared differences in long-baseline experiments such as T2HK and DUNE and reactor experiment such as JUNO can resolve the DLMA conundrum to better than $3\sigma$ within 1 year for T2HK and little more than 3 years for DUNE.
\end{abstract}
\maketitle

\section{Introduction}
Within the standard three-generation neutrino oscillation paradigm, the Large Mixing Angle (LMA) solution provides an excellent explanation to the solar neutrino problem \cite{Aharmim:2005gt} and has been independently tested and confirmed by the KamLAND \cite{Abe:2008aa} reactor antineutrino experiment.  The solar and KamLAND data jointly restrict the LMA solution to $\Delta m_{21}^2\simeq 7.5\times 10^{-5}$ eV$^2$ and $\sin^2\theta_{12} \simeq 0.3$ \cite{Esteban:2018azc}. In particular, the high significance compatibility of the solar neutrino data with resonant MSW matter effect inside the sun rules out all values of $\sin^2\theta_{12} > 0.5$ for $\Delta m_{21}^2 > 0$. However, this apparently robust conclusion runs into trouble \cite{Miranda:2004nb,Escrihuela:2009up} as soon as one allows for the presence of (large) Non-Standard neutrino Interactions (NSIs) \cite{Wolfenstein:1977ue}. It has been shown  that in presence of NSI we get two nearly degenerate solutions that fits the solar neutrino data -- the standard LMA solution ($\Delta m^2_{21}\simeq 7.5\times 10^{-5}$ eV$^2$, $\sin^2\theta_{12}\simeq 0.3$) mentioned above, and another solution at ($\Delta m^{2D}_{21}\simeq 7.5\times 10^{-5}$ eV$^2$, $\sin^2\theta_{12}^D\simeq 0.7$). This latter solution has been called the Dark LMA (DLMA) solution \cite{Miranda:2004nb} which has the same $\Delta m_{21}^2$ as in LMA solution but $\sin^2\theta_{12}^D=1-\sin^2\theta_{12}$. The conditions that need to be satisfied for the degeneracy to be exact include NSIs corresponding to scattering off both up and down type quarks with NSI couplings proportional to their respective charges \cite{Gonzalez-Garcia:2013usa,Bakhti:2014pva,Coloma:2016gei,Coloma:2017egw}. However even without this assumption, global analysis of solar neutrino and KamLAND data give the above mentioned approximately degenerate solutions at LMA and DLMA \cite{Esteban:2018ppq}. 

The existence of the DLMA solution not only has consequences for model building but has phenomenological implications as well. In particular, once both ``octants" of $\theta_{12}$ become viable, neutrino mass hierarchy determination at oscillation experiments becomes a problem owing to the  ``generalised mass hierarchy degeneracy'' \cite{Coloma:2016gei},wherein for pure vacuum oscillations, the neutrino mass matrix and hence the neutrino oscillation probabilities remain invariant under the simultaneous transformations $\Delta m_{31}^2 \leftrightarrow -\Delta m_{32}^2$, $\sin^2\theta_{12} \leftrightarrow \cos^2\theta_{12}$ and $\delta_{cp} \leftrightarrow \pi - \delta_{cp}$. This poses a serious challenge for experiments such as JUNO \cite{An:2015jdp,Choubey:2003qx} that are attempting to measure the sign of $\Delta m_{31}^2$ via vacuum oscillations \cite{Bakhti:2014pva}. Presence of matter effects could in principle break this degeneracy, however, if one adds the possibility of having NSI in the neutrino sector, then in addition to the transformations mentioned above if we also transform the NSI parameters $\epsilon_{ee} \rightarrow \epsilon_{ee} - 2$ and $\epsilon_{\alpha\beta} \rightarrow \epsilon_{\alpha\beta}^*$, then the generalised mass hierarchy degeneracy is restored and now even experiments such as DUNE \cite{Abi:2018dnh} that aim to determine the sign of $\ma$ via observation of earth matter effects are unable to do so \cite{Coloma:2016gei}. As long as this degeneracy remains unresolved, oscillation experiments will not be able to measure the sign of $\Delta m_{31}^2$, unless the issue of the octant of $\theta_{12}$ is resolved. Likewise, the octant of $\theta_{12}$ can be determined at oscillation experiments only if the sign of $\Delta m_{31}^2$ is known.   

Therefore, it is obvious that both cannot be determined simultaneously using oscillation experiments alone and one needs to look for inputs from non-oscillation experiments for the resolution of the LMA vs DLMA conundrum. This can be done in two ways. The first way is of course to independently constrain the NSI parameters themselves such that the values of the NSI parameters required for the DLMA solution are disfavored, disfavoring in turn the DLMA solution \cite{Coloma:2017egw,Coloma:2017ncl}. Constraints on NSI parameters have been obtained from neutrino scattering experiments. However, these constraints depend on the mass of the mediator particle. For heavy mediators, the best constraints come from the CHARM \cite{Dorenbosch:1986tb} and NuTeV experiments \cite{Zeller:2001hh}, while the current data from COHERENT \cite{Akimov:2018vzs} severely constrains the NSI parameters to mediator masses down to 10 MeV. However, there still remains allowed a narrow window of NSI parameters for which the DLMA solution is allowed \cite{Denton:2018xmq}. Indeed the global analysis \cite{Coloma:2019mbs} of neutrino oscillation data and COHERENT data allows the DLMA solution for certain (restricted) class of NSI models, with some cases even favoring the DLMA solution over the LMA solution (cf. Fig. 8 of \cite{Coloma:2019mbs}). Future data from CONUS and more data from COHERENT could narrow down this range severely \cite{Denton:2018xmq}, however, this remains to be seen, and it is possible that the DLMA solution could survive the constraints from neutrino scattering experiments for very low mediator masses and/or specific NSI models. Hence, neutrino scattering data may or may not be able to completely rule out the DLMA solution in a truly model independent way. 

The other way of resolving the $\theta_{12}$ vs $\theta_{12}^D$ conundrum is to determine the neutrino mass hierarchy in a non-oscillation experiment and then test $\theta_{12}$ vs $\theta_{12}^D$ in an oscillation experiment. In this work we study this second possibility, and show for the first time that it is possible for neutrino oscillation experiments to resolve the LMA vs DLMA degeneracy, once the mass hierarchy is determined using a different kind of physics. 
Such a non-oscillation experiment could be neutrino-less double beta decay, which depends crucially on the neutrino mass hierarchy. The implications of DLMA for neutrino-less double beta decay have been presented in \cite{N.:2019cot}. It was pointed out in that work that predictions for neutrino-less double beta decay for inverted hierarchy (IH) remain invariant under change from LMA to DLMA. However, for normal hierarchy (NH), the predicted value of the effective mass $m_{\beta\beta}$ shifts to higher values into the ''desert region", increasing the prospects of a positive signal in this very important class of experiments. Since the next generation neutrino-less double beta decay experiments are expected to cover the entire predicted range for IH, we will know the neutrino mass hierarchy in the near future, if neutrinos are Majorana particles. Here, we will study in detail the possibility of determining the octant of $\theta_{12}$ at long-baseline experiments such as T2HK \cite{Abe:2015zbg} or DUNE \cite{Abi:2018dnh} and show that owing to an intrinsic correlation between $\theta_{12}$ and $|\Delta m_{31}^2|$ in $P_{\mu\mu}$, it is impossible to say anything about $\theta_{12}$ at all from data from long-baseline experiments alone, even if the sign of $\Delta m_{31}^2$ was known. We will then show how we can break this correlation using data from a complementary oscillation experiment such as JUNO which measures $P_{ee}$. We will show that $P_{\mu\mu}$ measured at long-baseline experiments and $P_{ee}$ measured at reactor experiments together will be able to convincingly settle the issue of the octant of $\theta_{12}$ and resolve the DLMA conundrum.

The rest of the paper is organised as follows. In section \ref{sec:prob} we discuss the neutrino oscillation probabilities to show how the long-baseline experiments depend on the octant of $\theta_{12}$. In section \ref{sec:results} we discuss the correlation between $\theta_{12}$ and $|\Delta m_{31}^2|$ and the role of combining JUNO with T2HK (or DUNE) in breaking this correlation. In section \ref{sec:compare} we compare the sensitivity of T2HK vs DUNE in resolving the DLMA solution. Section \ref{sec:0nubb} discusses the role of neutrino-less  double beta decay experiments briefly. Finally, we end in section \ref{sec:conclusion} with our conclusions.

\section{LMA vs DLMA and Oscillation Probabilities \label{sec:prob}}

All results in this work are obtained within the exact three-generation paradigm, taking earth matter effect into account. However, the resolution of the octant of $\theta_{12}$ can be understood simply from the expressions of oscillation probabilities in vacuum. Therefore for simplicity, we begin by discussing oscillation probabilities in vacuum. 
In vacuum, the $\nu_\mu$ survival probability using the standard parametrisation of the neutrino mixing matrix is given as 
\begin{widetext}
\be
P_{\mu\mu} = 1 - P_{21}\sin^2\Bigg(\frac{\Delta m_{21}^2L}{4E} \Bigg )
&-& 4s_{23}^2c_{13}^2\bigg(c_{12}^2c_{23}^2 + s_{12}^2s_{23}^2s_{13}^2 - \frac{1}{2}\sin 2\theta_{12}\sin2\theta_{23}s_{13}^2\cos\delta\bigg)
\sin^2\Bigg(\frac{\Delta m_{31}^2L}{4E}\Bigg ) \nn \\
&-&  4s_{23}^2c_{13}^2\bigg(s_{12}^2c_{23}^2 + c_{12}^2s_{23}^2s_{13}^2 - \frac{1}{2}\sin 2\theta_{12}\sin2\theta_{23}s_{13}^2\cos\delta\bigg)
\sin^2\Bigg(\frac{\Delta m_{32}^2L}{4E} \Bigg )
\,,
\label{eq:pmmfull}
\ee
\end{widetext}
where $\Delta m_{ij}^2=m_i^2-m_j^2$, $s^2_{ij}$ and $c^2_{ij}$ imply $\sin^2\theta_{ij}$ and $\cos^2\theta_{ij}$, respectively, while $E$ and $L$ are the neutrino energy and distance travelled. $P_{21}$ associated with the $\Delta m_{21}^2$-driven oscillatory term is symmetric with respect to the interchange $\sin^2\theta_{12} \leftrightarrow \cos^2\theta_{12}$ and hence not interesting to us. The expression in Eq.~(\ref{eq:pmmfull}) shows that  $P_{\mu\mu}$ is invariant under simultaneous exchange of $\sin^2\theta_{12} \leftrightarrow \cos^2\theta_{12}$ and $|\Delta m_{31}^2| \leftrightarrow |\Delta m_{32}^2|$. Therefore, if the goal is to distinguish $\sin^2\theta_{12}$ from $\cos^2\theta_{12}$ ($=\sin^2\theta_{12}^D$), the experiment should be able to distinguish the oscillatory frequency corresponding to $|\Delta m_{31}^2|$ from the frequency corresponding to $|\Delta m_{32}^2|$. The difference in $P_{\mu\mu}$ between $\sin^2\theta_{12}$ and $\sin^2\theta_{12}^D=\cos^2\theta_{12}$ is given as, 
\be
\Delta P_{\mu\mu} = &&4s_{23}^2c_{13}^2\cos2\theta_{12}\bigg(c_{23}^2 - s_{23}^2s_{13}^2\bigg)\nn\\
&&\sin\Bigg(\frac{\Delta m_{21}^2L}{4E} \Bigg )
\sin\Bigg(\frac{(2\Delta m_{31}-\Delta m_{21}^2)L}{4E} \Bigg )
\,.
\label{eq:delpmm}
\ee

\begin{figure}
\includegraphics[width=0.45\textwidth]{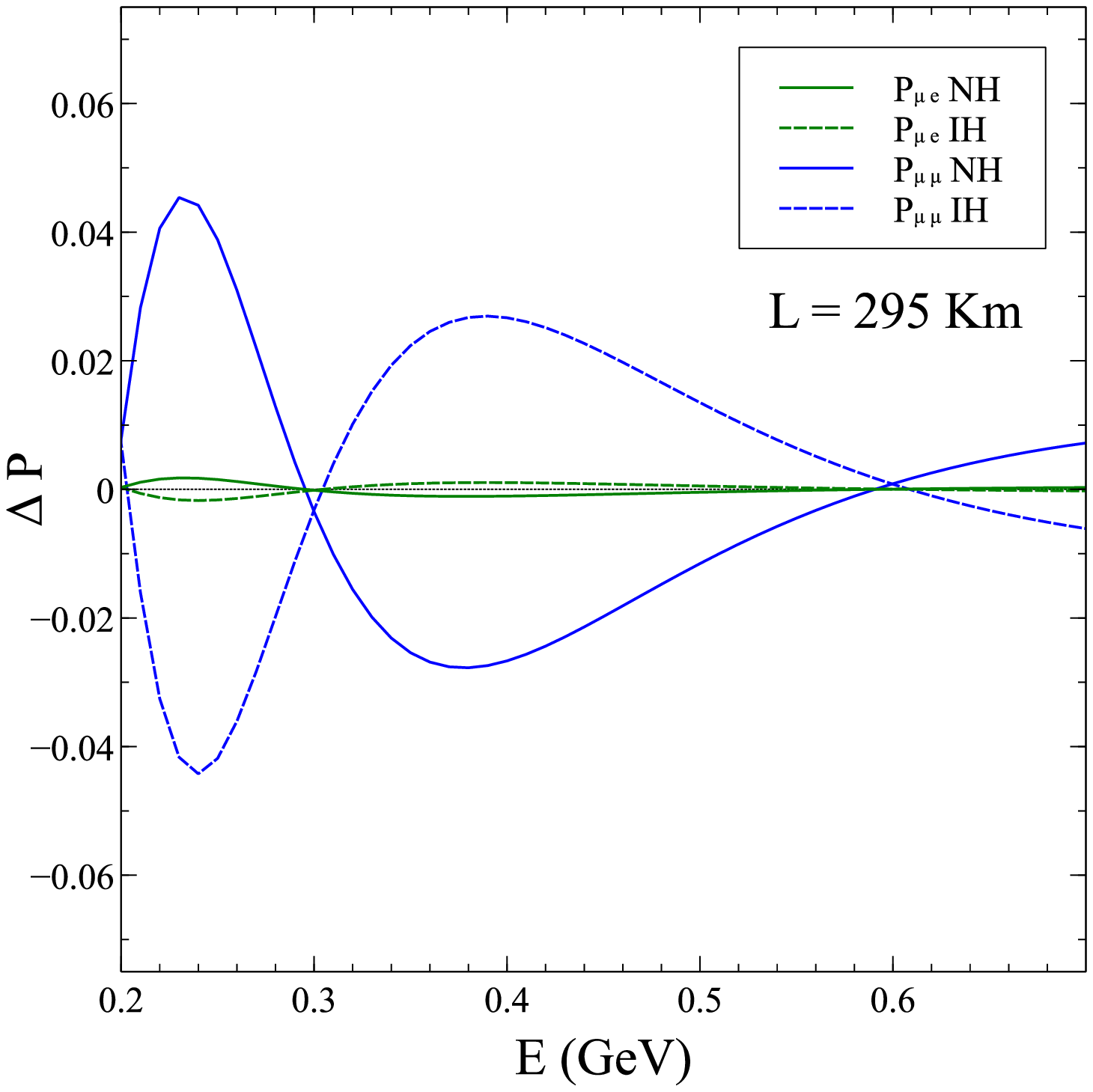}
\caption{\label{fig:prob_diff1} Difference in probability corresponding to $\theta_{12}$ and $\theta_{12}^D$, as a function of energy. The plot is shown for $L=295$ km and	without matter effect. The black lines show $\Delta P_{\mu\mu}$ while the green lines show $\Delta P_{\mu e}$. The solid lines are for NH while the dashed lines are for IH. }
\end{figure}

\noindent
Note that this difference goes to zero close to the oscillation maximum corresponding to $\ma$, and is extremal when $\sin(2\Delta m_{31}-\Delta m_{21}^2)L/4E \simeq 1$. The difference will also go to zero if $c_{23}^2 = s_{23}^2s_{13}^2$, which is safely avoided for the current allowed values of $\theta_{13}$ and $\theta_{23}$.  The corresponding difference in the $P_{\mu e}$ channel between $\sin^2\theta_{12}$ and $\sin^2\theta_{12}^D$ is  
\be
\Delta P_{\mu e} = &&s_{23}^2\sin^22\theta_{13}\cos2\theta_{12}\nn\\
&&\sin\Bigg(\frac{\Delta m_{21}^2L}{4E} \Bigg )
\sin\Bigg(\frac{(2\Delta m_{31}-\Delta m_{21}^2)L}{4E} \Bigg )
\,.
\label{eq:delpme}
\ee
In Fig.~\ref{fig:prob_diff1} we show $\Delta P_{\mu\mu}$ and $\Delta P_{\mu e}$ as a function of energy for the T2HK baseline of $L=295$ km for normal (NH) and inverted (IH) mass hierarchy. We define NH and IH corresponding to $\Delta m_{31}^2>0$  and $\Delta m_{31}^2<0$, respectively. Here and throughout the rest of the paper we take true values of the parameters as follows: $\Delta m_{21}^2=7.5 \times 10^{-5}$ eV$^2$,  $\Delta m_{31}^2=2.52 \times 10^{-3}$ eV$^2$, $\sin^2\theta_{12}=0.31$, $\theta_{12} = 8.46^{\circ}$, $\theta_{23}=42^{\circ}$ and $\delta_{cp}=-90^\circ$. For a given mass hierarchy, extrema in $\Delta P_{\mu\mu}$ and $\Delta P_{e\mu}$ appear close to $E\sim \Delta m_{31}^2L/(2n+1)\pi$, with $n=2,$ 1, 0. We see a maxima in $\Delta P_{\mu\mu}$ and $\Delta P_{e\mu}$ for $n=2$, minima for $n=1$ and back again a maxima for $n=1$, as was expected from Eqs.~(\ref{eq:delpmm}) and (\ref{eq:delpme}). This flipping of $\Delta P_{\mu\mu}$ with neutrino energy  is also a signal for the octant of $\theta_{12}$. The presence of $\Delta m_{21}^2$ has two-fold effect. Firstly, it dampens $\Delta P$ as we go to higher energies.  This comes from the first oscillatory term $\sin(\Delta m_{21}^2 L/4E)$ in Eqs.~(\ref{eq:delpmm}) and (\ref{eq:delpme}). Since this term decreases as we increase energy, the damping is seen to increase with energy in Fig.~\ref{fig:prob_diff1} . Secondly, it determines the exact position of the $\Delta P$ extrema according to the second oscillatory term. 

A comparison of Eqs.~(\ref{eq:delpmm}) and (\ref{eq:delpme}) shows that $\Delta P_{\mu e}/\Delta P_{\mu \mu} = s_{13}^2/(\cos2\theta_{23}+s_{23}^2c_{13}^2)$, which for maximal $\theta_{23}$ gives $\Delta P_{\mu e}/\Delta P_{\mu \mu} = 2s_{13}^2/c_{13}^2$. Thus, for the measured value of $\theta_{13}$, $\Delta P_{\mu e}$ is suppressed by two orders of magnitude compared to $\Delta P_{\mu\mu}$. This difference can be seen in Fig.~\ref{fig:prob_diff1} and hence, we conclude that it is the $P_{\mu\mu}$ (disappearance) channel which will bring in the sensitivity to the octant of $\theta_{12}$, while the role of the $P_{\mu e}$ (appearance) channel will be seen to be minimal.

\begin{figure}
\includegraphics[width=0.45\textwidth]{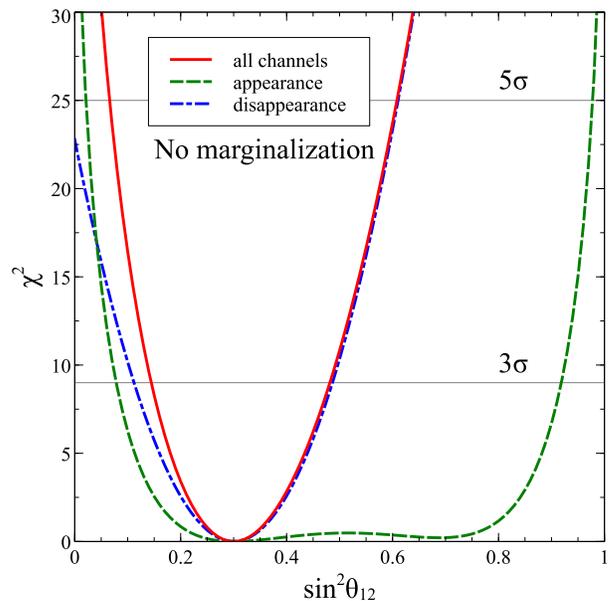}
\caption{\label{fig:chi_nomarg}$\chi^2$ as a function of $\sss$ for T2HK mock data generated at $\sss=0.31$ corresponding to the LMA solution and fitted by varying $\sin^2\theta_{12}$ while all other parameters are fixed. Results are shown for analysis of appearance channel data only, disappearance channel data only and for the combination of both appearance and disappearance channel data. }
\end{figure}

We show in Fig.~\ref{fig:chi_nomarg} the $\chi^2$ and hence the significance with which T2HK will be able to distinguish LMA from DLMA, if all oscillation parameters are kept fixed at their assumed true value in the fit. We have taken the experimental settings for T2HK as given in \cite{Abe:2015zbg} and considered the mock data corresponding to running the experiment for 2.5+7.5 years in $\nu$+$\bar\nu$ mode, respectively. The mock data is generated at $\sss=0.31$ and we want to rule out the higher octant of $\theta_{12}$ in general and the DLMA solution in particular. Shown are the  $\chi^2$ obtained from the analysis of disappearance data, appearance data, and both data sets combined. The appearance channel on its own is unable to resolve the degeneracy. However, it puts a weak lower and upper limit on $\sss$, as $P_{\mu e}$ depends on $\sin^22\theta_{12}$. We see that the disappearance data alone can rule out the DLMA solution at more than 5$\sigma$, if all the other parameters were known. However, the other oscillation parameters are not completely known, and hence should be allowed to vary to account for correlations between them. Indeed there is a correlation between $\Delta m_{31}^2$ and $\theta_{12}$, which we discuss next.

\section{Mass Squared Difference and $\theta_{12}$ \label{sec:results}}

\begin{figure}
\includegraphics[width=0.45\textwidth]{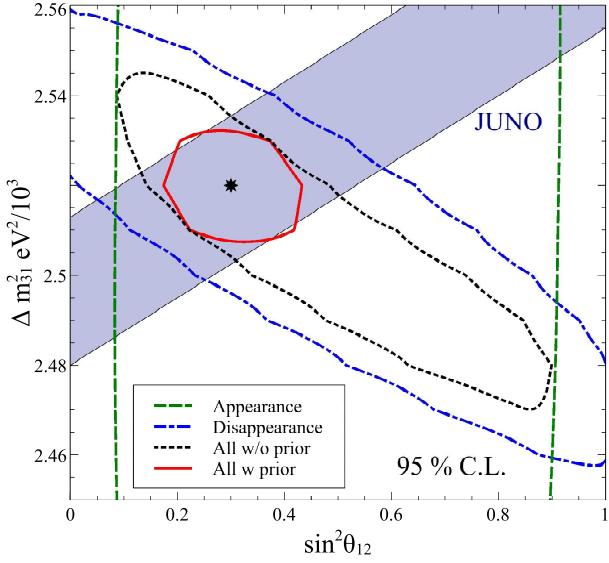}
\caption{\label{fig:chi_cont}The 95\% C.L. expected allowed areas in the $\ma-\sss$ plane for T2HK. The blue dot-dashed contour is obtained using T2HK disappearance channel only, the green dashed curve using T2HK appearance channel only and the black dotted curve using both. The blue shaded area is the expected 95\% allowed area from JUNO. The solid red contour is obtained when we apply the prior form JUNO to the analysis of projected T2HK data. }
\end{figure}

Fig.~\ref{fig:chi_cont} presents the 95\% C.L. allowed area in the $\Delta m_{31}^2 - \sin^2\theta_{12}$ plane expected at T2HK from disappearance channel (blue dot-dashed curve), appearance channel (green dashed curve) and by combining the two (black dotted curve). The contours are marginalised over $\theta_{23}$, $\theta_{13}$ and $\dcp$.  The assumed true value in this parameter space is shown by the black star. We find that once we allow $\ma$ to vary, the constraint on $\sss$ is seen to be completely lost. The appearance channel puts only a weak constraint on $\sin^22\theta_{12}$, as was the case in Fig.~\ref{fig:chi_nomarg}. However, for the disappearance channel the effect is drastic and we find that nearly the entire $\sss$ range gets allowed. This happens because of the anti-correlation between $\ma$ and $\sss$ in the $\numu$ survival probability which can be explained easily if we rewrite Eq.~(\ref{eq:pmmfull}) in terms of an effective mass squared difference $\Delta m^2_{\mu\mu}$ such that \cite{deGouvea:2005hk,Minakata:2006gq,Zhan:2008id}
\be
P_{\mu\mu} \simeq 1-4\sin^2\theta_{23}\cos^2\theta_{13}(1-\sin^2\theta_{23}\cos^2\theta_{13})\sin^2\Bigg(\frac{\Delta m_{\mu\mu}^2L}{4E} \Bigg )  \nn
\,,
\label{eq:pmmmeff}
\ee
\be
\!\!\!\!\!\!
\Delta m_{\mu\mu}^2\!=\!\ma \!-\! \ms(\cos^2\theta_{12}-\cos\delta\sin\theta_{13}\sin2\theta_{12}\tan\theta_{23})
\label{eq:delmmmeff}
\ee
The experiment measures the effective parameter $\Delta m_{\mu\mu}^2$ which depends on $\ma$ and $\sin^2\theta_{12}$ in an anti-correlated way and hence $P_{\mu\mu}$ alone will never be able to measure $\sss$ in a long-baseline experiment. A given measured value of $\Delta m_{\mu\mu}^2$ can always be reproduced for lower $\ma$ and higher $\sss$, and/or vice-versa. Adding the $P_{\mu e}$ channel to the analysis along with the $P_{\mu\mu}$ channel (shown by black short-dashed line) improves the constraint on $\sss$ only marginally on the two edges of the scale, as was expected from the green-dashed lines in Figs.~\ref{fig:chi_nomarg}. However, this is unable to break the $\ma$ and $\sss$ anti-correlation coming from $\Delta m_{\mu\mu}^2$ and as a result the DLMA solution continues to get allowed. In order to remedy this, we need an independent measurement of $\ma$ which can break the $\ma$ and $\sin^2\theta_{12}$ anti-correlation in the $P_{\mu\mu}$ channel. We use here the expected constraints from the JUNO \cite{An:2015jdp} experiment which will observe $\bar\nu_e$ disappearance probability $P_{ee}$ and will measure the effective mass squared difference \cite{deGouvea:2005hk,Minakata:2006gq,Zhan:2008id}
\be
\Delta m^2_{ee} = \ma - \ms\sss
\label{eq:mee}
\,.
\ee
Note that the dependence of $\Delta m^2_{ee}$ on $\theta_{12}$ is complementary to what we had for $\Delta m^2_{\mu\mu}$. While the former depends on $\sss$, the latter on $\cos^2\theta_{12}$. We show in the blue band the 95\% C.L. expected allowed area in the $\ma-\sss$ plane from JUNO \cite{An:2015jdp} and find that $\ma$ and $\sss$ are correlated for JUNO, while they are anti-correlated for T2HK.  Including the JUNO constraint as a prior in our analysis of the T2HK data gives us the contour shown by the red solid curve. The degeneracy between $\ma$ and $\sss$ is broken, giving us a unique solution in $\sss$ and the DLMA solution can now be completely ruled out.


\section{Comparing T2HK with DUNE \label{sec:compare}}

\begin{figure}
\includegraphics[width=0.45\textwidth]{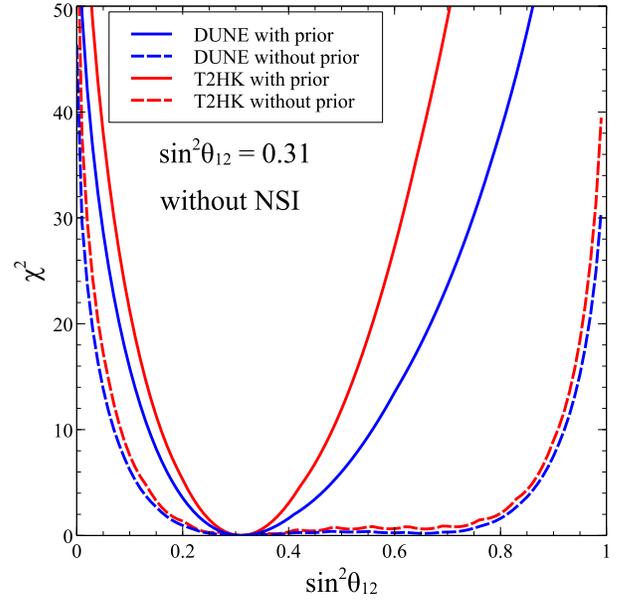}
\caption{\label{fig:chi}$\chi^2$ as a function of $\sss$ for mock data generated at 
$\sss=0.31$. The parameters $\theta_{23}$, $\theta_{13}$, $\dcp$ as well as $|\ma|$ are allowed to vary freely in the fit. Shown are the sensitivity expected at T2HK as well as DUNE with and without prior from JUNO. }
\end{figure}

In Fig.~\ref{fig:chi} we show the $\chi^2$ as a function of $\sss$ for mock data generated at 
$\sss=0.31$. We have shown here the expected $\chi^2$ from combined appearance and disappearance data at T2HK (red lines) as well as DUNE (blue lines). For DUNE we have taken experimental specifications as in \cite{Abi:2018dnh} and 10 years of running. The oscillation parameters $\theta_{23}$, $\theta_{13}$, $\dcp$ as well as $|\ma|$ are varied freely in the fit. Results are shown without taking the prior on $\Delta m_{ee}^2$ from JUNO (dashed lines) as well as when the JUNO prior in included (solid lines). As in Fig.~\ref{fig:chi_cont}, we see that without the prior on $\Delta m^2_{ee}$ from JUNO, it is impossible for the long-baseline experiments to put any meaningful constraint on $\sss$. However, once the information from JUNO is included, the disappearance channel at these experiments can rule out the DLMA solution at more than 5$\sigma$ significance at T2HK and more than 4$\sigma$ at DUNE.
 
A final comment on the statistics needed at these experiments to rule out DLMA is in order. We find that if we keep the $\nu$ to $\bar\nu$ fraction at these experiments same as in their proposal, we could rule out DLMA at $3\sigma$ C.L. by combining JUNO data with 1 year of T2HK data and slightly more than 3 years of DUNE data.

\section{Role of Neutrino-less Double Beta Decay \label{sec:0nubb}}

\begin{figure}
\includegraphics[width=0.45\textwidth]{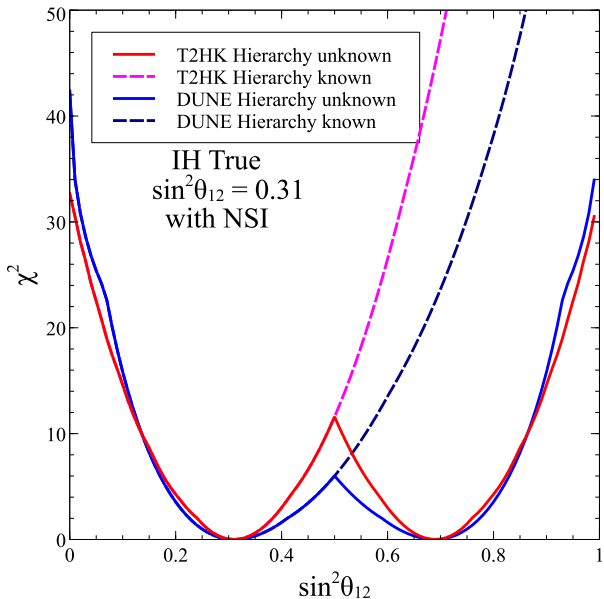}
\caption{\label{fig:mheffect} The impact of the neutrino mass hierarchy on the determination of the octant of $\theta_{12}$.}

\end{figure}
So long we had fixed the neutrino mass hierarchy, assuming that it will be determined by a non-oscillation experiment. Neutrino-less double beta decay should be able to fix the neutrino mass hierarchy \cite{Furry:1939qr,DellOro:2016tmg}. Information on neutrino mass hierarchy could also come from future cosmological data \cite{RoyChoudhury:2019hls,Vagnozzi:2017ovm}.  
In Fig.~\ref{fig:mheffect}, we show the impact of neutrino mass hierarchy on ruling out the DLMA solution. We present two cases each for T2HK and DUNE - the solid lines show the $\chi^2$ when the hierarchy is unknown (red for T2HK and blue for DUNE) while the dashed lines show the $\chi^2$ when the hierarchy is known (magenta for T2HK and dark-blue for DUNE). If the neutrino mass hierarchy is known, we keep the mass hierarchy fixed in our fit while if it is unknown we marginalise our $\chi^{2}$ over mass hierarchy. The data is generated using oscillation parameters given in section \ref{sec:prob} and with no NSI, while in the fit we vary both oscillation parameters as well as NSI parameters within their allowed ranges.  We see that due to the generalised mass hierarchy  degeneracy, for the hierarchy unknown case, the fit in the presence of NSI parameters gives two degenerate solutions corresponding to the LMA (true solution) and DLMA (fake solution). On the other hand, if we know the mass hierarchy through measurements at neutrino-less double beta decay experiments and/or cosmology, the corresponding $\chi^2$ show that the DLMA solution can be disfavoured at a very high significance. We also note that the entire higher octant corresponding to $\sss > 0.5$ can be ruled out at greater than $3\sigma$ by data from T2HK. For DUNE the significance is slightly less, but still competitive.

\section{Conclusions \label{sec:conclusion}}

In presence of NSIs the solar neutrino data is known to give two (nearly) degenerate solutions with same $\ms$ but with $\sss\simeq 0.3$ and $\sin^2\theta_{12}^D\simeq 0.7$, known as the LMA and DLMA solutions, respectively. The existence of this degeneracy has both model building as well as  phenomenological implications. One being the so-called generalised mass hierarchy degeneracy wherein the neutrino mass matrix in vacuum remains invariant under the simultaneous exchange of $\Delta m_{31}^2 \leftrightarrow -\Delta m_{32}^2$, $\sin^2\theta_{12} \leftrightarrow \cos^2\theta_{12}$ and $\delta_{cp} \leftrightarrow \pi - \delta_{cp}$. This is a problem for experiments like JUNO which aim to measure the neutrino mass hierarchy by observing neutrino oscillations in vacuum. While presence of matter effect could resolve this degeneracy for standard neutrinos, in presence of NSI the transformations $\epsilon_{ee} \rightarrow \epsilon_{ee} - 2$ and $\epsilon_{\alpha\beta} \rightarrow \epsilon_{\alpha\beta}^*$ are allowed and hence the generalised mass hierarchy degeneracy is restored, making simultaneous measurement of neutrino mass hierarchy and octant of $\theta_{12}$ impossible at neutrino oscillation experiments alone. 

One way to resolve this problem is to constrain NSI at neutrino scattering experiments. In this work we showed that it is possible to resolve this problem and measure the octant of $\theta_{12}$ in neutrino oscillation experiments if we determine the neutrino mass hierarchy at a non-oscillation experiment such as neutrino-less double beta decay or cosmology. We showed the importance of the $P_{\mu\mu}$ channel in long-baseline experiments in distinguishing LMA from DLMA. We next showed the anti-correlation between $\ma$ and $\sss$ in $P_{\mu\mu}$ since what is measured in the $P_{\mu\mu}$ channel at long-baseline experiments is the combination $\Delta m_{\mu\mu}^2$. We then pointed out that JUNO measures the combination $\Delta m_{ee}^2$ via the  $P_{ee}$ disappearance channel. We showed that since for $\Delta m_{ee}^2$ the parameters $\ma$ and $\sss$ are correlated instead of being anti-correlation as in $P_{\mu\mu}$, a combination of these two disappearance channel can unambiguously pin down the octant of $\theta_{12}$ and hence can rule out the DLMA solution, if the mass hierarchy was determined in neutrino-less double beta decay and/or cosmology.  From a statistical analysis of mock data and taking the prior from JUNO, we showed that the DLMA solution can be ruled out at T2HK and DUNE at $5\sigma$ and $4\sigma$ C.L., respectively.  Finally, we found that the DLMA solution can be ruled out at $3\sigma$ by combining JUNO data with just 1 year of T2HK running and little more than 3 years of DUNE running.

\section{Acknowledgment}
We acknowledge support from HRI where major part of this work was completed. 
The HRI cluster
computing facility (http://cluster.hri.res.in) is also acknowledged. We thank S. Goswami for 
her valuable inputs and suggestions.  
This project has received funding from the European Union's Horizon
2020 research and innovation programme InvisiblesPlus RISE
under the Marie Sklodowska-Curie
grant  agreement  No  690575. This  project  has
received  funding  from  the  European
Union's Horizon  2020  research  and  innovation
programme  Elusives  ITN  under  the 
Marie  Sklodowska-Curie grant agreement No 674896.

\bibliography{ref}

\end{document}